\documentstyle[12pt]{article}
\begin{document}
\baselineskip .3in
\begin{titlepage}
\begin{center}{\large{\bf The dimensional properties of the hadron}}
\vskip .2in S.Mukherjee(Banerjee)$^{\dag}$ \\
 \vskip .1in Dublin
Institute of Technology,Dublin,Ireland. \vskip .1in
S. N. Banerjee  \\
\vskip .1in
Department of Physics, Jadavpur University \\
Calcutta 700032, India.\\

\end{center}

\vskip .3in \noindent {\bf Abstract} \vskip .1in \noindent The
statistical model of the hadron as a quarkonium system suggests
the dimensional dependence of some of the properties of the hadron
like the scaling in its probability density with the anomalous
dimension as the exponent,the universality,the power law scaling
in transverse momentum in hadron production.Recent experimental
findings are founding to be in agreement with the corresponding
predictions of the model.


\vskip .3in

\noindent $^{\dag}$ E-mail:sarbarimukherjee09@gmail.com

\end{titlepage}

\newpage
 \vskip .1in \noindent
{\bf 1.Introduction} \vskip .1in \noindent The statistical model
of the hadron as a quarkonium system,proposed almost three decades
ago[1],has been widely used to study the properties of hadrons.The
model,however,has undergone modifications along with its various
ramifications and applications[2,3] since its very inception in
1981.The probability density $p(\xi)$ of the hadron runs as
\begin{equation}
p(\xi)=A\xi^{3\mu/2}\Theta(\xi)
\end{equation}
where $\xi= r-r_{0}$,$\Theta $ is the usual step function,$r_{0}$
is a parameter corresponding to the size of the meson and $A=[4\pi
r_{0}^{\frac{3\mu}{2} +3},B(3,\frac{3\mu}{2}+1)]^{-1}$ where B is
the Beta function.The parameter $ \mu $ becomes equal to 3.66 or 1
according as the exchange effect is taken into consideration or
ignored[2].In other words,we come across the
asymptotic(fractal)dimension $D=9/2$ at high energy or $D=8.5$ at
comparatively moderate energy.As
\begin{equation}
p(\xi)\sim
(\xi/r_{0})^{D-D_{T}}r_{0}^{-3}=(\xi/r_{0})^{D_{A}}r_{0}^{-3}
\end{equation}
where $D_{T}=3$ is the topological dimension,$D_{A}$,the anomalous
dimension,appearing as an exponent in our subsequent discussion is
either 1.5 or 5.5. \vskip .1in \noindent {\bf 2.Properties of the
hadron} \vskip .1in \noindent It is a remarakable feature of the
model that $ p(\xi)$ for $r< r_{0}$ is non-analytic because of the
branch cut and it scales as $\xi^{D_{A}}$ with $D_{A}$,the
non-integer exponent.As it is a scaling function,we have
\begin{equation}
p(\lambda\xi)\sim \lambda^{D_{A}}p(\xi)
\end{equation}
where $\lambda $ is the scale factor and if $\lambda\xi =\eta$ for
all positive values of $ \lambda$,we get
\begin{equation}
p(\eta) \sim \lambda^{D_{A}}p(\eta/\lambda)
\end{equation}
This type of symmetry helps us to connect physical phenomena at
different length scales.Since the l.h.s. of (4) is independent of
$ \lambda $,$p(\eta)$ must be a homogenous function and the only
possible form of $p(\eta)$ is the power law $p(\eta)\sim
\eta^{D_{A}}$.In the case of geometric scaling,the scaling laws
with the anomalous exponents arise in systems that are described
by fractals and that random structures are known to display
fractal properties in terms of the anomalous exponent.\vskip .1in

The dimensionality of the hadron as a fractal becomes a
quantitative measure of the qualitative property of a strucuture
that is self-similar.Thus $p(\xi)$ of the hadron displaying
self-similarity,suggests a structure that looks the same at
different length scales.Hence we come across the universality
property suggesting that many different systems show the same
critical behaviour.The apparent universality property of the
hadron,because of the same $D_{A}$ carries the meaning that all
hadrons behave in the same manner close to the critical point as
the static critical exponents become identical.In other
words,$D_{A}$ becomes the universal parameter and different
hadrons display the same critical behaviour.Gazdzicki et al[4]
have postulated that the energy spectrum of the hadronising matter
obeys a power law distribution and that the transverse mass
storage of $\Pi^{0}$ mesons for $m_{t}> 1 Gev/c^{2}$ obeys the law
${c{(m_{t}/\Lambda)}}^{-a}$, similar to the hadronic multiplicity
where the transverse mass $m_{t}$ is $\sqrt{m^{2}+p_{t}^{2}}$
where $p_{t}$ is the transverse momentum.It results in a similar
power law in mass distribution i.e. $\rho(m) =cm^{-a}$ where the
normalisation constant c and the parameter a are the same for
different hadron species.It describes the yields of neutral mesons
from eta to upsilon i.e. for  $m=0.5$ to 10 $Gev/c^{2}$ with $a
\sim 8-10$.We would like to assert that the probability of
production of hadrons of transverse mass $m_{t}$ would be
dominated in the neighbourhood of $\xi = \bar{\xi}$ such that
$\bar{\xi}$ is $m_{t}^{-1}$ and $r_{0}$ is the scale parameter.As
the probability for the multiplicity distribution of N-particles
obeys the law $\sum_{N=0}^{\infty}p_{N}=1$ and the average
multiciplicity $<N> = \sum_{N=0}^{\infty}p_{N}N $,as in [5], our
assertion is that $p_{N}(m_{t})$ is proportional to
$p(\bar{\xi},r_{0})$V where V is the volume of the hadron .Hence
$<N>$ is proportional to $(\bar{\xi}/r_{0})^{D_{A}}$ i.e. to
$(m_{t}/\Lambda)^{-D_{A}}$.Therefore,we get the transverse mass
distribution $\rho(m_{t})$ as $\rho(m_{t}) \propto
\frac{dN}{m_{t}^{2}dm_{t}}\sim (m_{t}/\Lambda)^{-D}$.\vskip .1in
\noindent \vskip .1in \noindent {\bf 3.Conclusion}\vskip .1in
\noindent As discussed before,for moderate energies $D=8.5$ and we
come across $\rho(m_{t})\sim (\frac{m_{t}}{\Lambda})^{-8.5}$
signalling close agreement with the corresponding experimental
findings [6].Further,it is also relevant and interesting to assert
that in the case of pure,high energy hadron collisions,inelastic
production processes of hadrons with high $p_{t}$ suggest
$p_{t}^{-4}$ type of behaviour i.e. the canonical asymptotic law
in the inclusive cross section.Incidentally,it turns out as a
close approximation to our prediction $p_{t}^{-D}$ where $D=4.5$.
corresponding to the asymptotic (fractal) dimension of hadron.This
power law prediction along with the reasonable value of the
exponent D clearly subscribes to the realistic view of the model
for the description of nature.

\newpage
\vskip .2in \noindent {\bf References}

\noindent[1]S.N.Banerjee et. al., Had.
J.4,203(1981);(E)Had.J.5,2157(1982); Had.J.6
440,760(1983);J.Phys.G.8,L61(1982);Cand.J.Phys.
61,532(1983);
Phys.Scr.27,163(1983);Ann.Phys.(N.Y.)150,150(1983);
Nuo.Cim.91A,247(1986);Acta Phys.Polon.B19,1011(1988).

\noindent [2]S.N. Banerjee et.al.,Had. J.
11,243(1989);12,179(1989);13,750(1990)

\noindent [3] S.N.Banerjee et. al.Int. J.Mod. Phys. A16,201(2001);
Int.J.Mod.Phys.A17,4939(2002); 13,750(1990);
Nucl.Phys.B.Proc.Suppl.142,13(2005);Mod.Phys.Lett.A24,509(2009).

\noindent [4]M.Gazdicki et al,Phys. Lett.B517,250(2001).

\noindent [5]I. Dremin et al, Phys.Rep.349,301(2001).

\noindent [6] S.N.Banerjee et al, Phys.Lett.B644,45(2007).

\end{document}